\documentclass[12pt]{article}
\usepackage{epsfig}
\usepackage{amssymb,amsmath}
\usepackage{setspace}
\pdfoutput=1
\usepackage{subfigure}
\usepackage{amssymb,amsmath}
\usepackage{graphicx}
\usepackage{color}
\usepackage{cancel}
\usepackage[colorlinks=true
,urlcolor=blue
,citecolor=blue
,linkcolor=blue
,pagecolor=blue
,linktocpage=true
,pdfproducer=medialab
]{hyperref}
\def\bi{\begin{itemize}}
\def\ei{\end{itemize}}

\def\to{\rightarrow}

\def\alt{\lesssim}
\def\agt{\gtrsim}

\usepackage{amsmath}

\newcommand{\bea}{\begin{eqnarray}}
\newcommand{\eea}{\end{eqnarray}}

\newcommand{\beq}{\begin{equation}}
\newcommand{\eeq}{\end{equation}}

 \makeatletter
\def\alt{\mathrel{\mathpalette\gl@align<}}
\def\agt{\mathrel{\mathpalette\gl@align>}}
\def\gl@align#1#2{\lower.6ex\vbox{\baselineskip\z@skip\lineskip\z@
\ialign{$\m@th#1\hfil##\hfil$\crcr#2\crcr\sim\crcr}}} \makeatother

\usepackage{longtable}

\begin{document}

\vspace*{1.0cm}

\begin{center}
\baselineskip 20pt {\Large\bf
A Promising Interpretation of Diphoton Resonance at 750 GeV
}
\vspace{1cm}

{\large
Xiao-Jun Bi$^{a,}$
Ran Ding$^{b,}$,
Yizhou Fan$^{c,}$,
Li Huang$^{c,}$,
Chuang Li$^{c,}$, \\
Tianjun Li$^{c,d,}$, 
Shabbar Raza$^{c,}$,
Xiao-Chuan Wang$^{e,}$,
Bin Zhu$^{f,}$
} \vspace{.5cm}

{\baselineskip 20pt \it $^a$
Key Laboratory of Particle Astrophysics, Institute of High Energy Physics,
Chinese Academy of Sciences, Beijing 100049, P. R. China
\\
{\it $^b$
Center for High-Energy
Physics, Peking University, Beijing, 100871, P. R. China
}
\\
{\it $^c$ 
State Key Laboratory of Theoretical Physics and Kavli Institute for Theoretical Physics China (KITPC),
Institute of Theoretical Physics, Chinese Academy of Sciences, Beijing 100190, P. R. China 
}
\\
{\it $^d$
School of Physical Electronics, University of Electronic Science and Technology of China,\\
Chengdu 610054, P. R. China 
}
\\
{\it $^e$
Department of Physics, Henan Normal University, Xinxiang, Henan, 453007, P.R.China
}
\\
{\it $^f$
Institute of Physics, Chinese Academy of sciences, Beijing 100190, P. R. China
}
}

\vspace{.5cm}

\vspace{1.5cm} {\bf Abstract}
\end{center}
Recently, an excess of events in diphoton channel with invariant mass of about 750 GeV has been reported by the ATLAS and CMS Collaborations. Considering it as a tantalizing hint for 
 new physics beyond the Standard Model (SM), we propose a simple extension of 
 the SM with an additional doublet Higgs $H^{'}$ and a singlet $s$. We consider the
 neutral component $H^{\prime }_0$ of $H'$ as the 750 GeV resonance, and assume that
 $s$ is lighter than 2.6 GeV. In particular,  $H^{\prime }_0$ can be produced 
 at tree level via $q{\bar q}$ production, and decay into a pair of $s$ at tree level.
 And then $s$ can decay into a pair of collimated photons, which cannot be distinguished
 at the LHC. We show that the diphoton production cross section can be from
 3 to 13 ${\rm fb}$, the decay width of $H^{\prime }_0$ can be from 30 to
 60 GeV, and all the current experimental constraints including dijet constraint
 can be satisfied.

\thispagestyle{empty}

\newpage

\addtocounter{page}{-1}

\baselineskip 18pt

\section{Introduction}
At the Run 2 of the Large Hadron Collider (LHC) with a center of mass energy of $\sqrt{s}=$ 13 TeV, both the
ATLAS~\cite{bib:ATLAS_diphoton} and CMS~\cite{bib:CMS_diphoton} Collaborations have reported an excess in diphoton channel with 
invariant mass of about 750 GeV. With an integrated luminosity of 3.2 ${\rm fb}^{-1}$,
the ATLAS Collaboration has observed a local $3.6\sigma$ 
 excess at the diphoton invariant mass around 747~GeV, assuming 
 a narrow width resonance. For a wider width resonance, the signal significance 
 increases to $3.9\sigma$ with a preferred width about 45~GeV.
Using 3.2 ${\rm fb}^{-1}$ of data, the CMS Collaboration found a diphoton excess
with a local significance of $2.6\sigma$ at invariant mass around 760 GeV. Assuming
the decay width around 45~GeV, the significance reduces to $2\sigma$. 
The excesses in the cross sections can be roughly estimated as
$\sigma_{pp\to \gamma \gamma}^{13~ {\rm TeV}} \sim 3-13~{\rm fb}$~~\cite{bib:ATLAS_diphoton, bib:CMS_diphoton}.
 It is interesting to note that the
CMS Collaboration did search for diphoton resonance~\cite{Khachatryan:2015qba} at $\sqrt{s} =$ 8 TeV and observed a slight excess $\sim$ 2$\sigma$ at an invariant mass of about 750 GeV but on the other hand the ATLAS Collaboration
 did not go beyond the mass of 600 GeV for this channel~\cite{Aad:2014ioa}. This indicates that the present ATLAS and CMS
observations at $\sqrt{s} =$ 13 TeV are consistent with their results at $\sqrt{s} =$ 8 TeV for diphoton channel.

Taking these results optimistically, the excess in 
diphoton channels may turn out to be the rays of light showing the dawn of the new era of long awaited physics beyond the Standard Model (SM).
In this study we interpret the excess of diphoton events as a hint for 
new physics beyond the SM.
The observed resonance can be naively understood as a spin-$0$ or $2$ particle with mass $750$ GeV because of the
Landau-Yang theorem~\cite{Landau:1948kw,Yang:1950rg},
Spurred by these interesting developments, new studies in model building for both effective and renormalizable frameworks~\cite{Hall:2015xds, Chala:2015cev, Angelescu:2015uiz,Gupta:2015zzs,Altmannshofer:2015xfo, Ding:2015rxx, Agrawal:2015dbf} have been carried out. Though apparently spin-$1$ resonance seems to be forbidden due to
Landau-Yang theorem~\cite{Landau:1948kw,Yang:1950rg},  it was shown in ~\cite{Chala:2015cev} that one can still trick it to obtain the diphoton excess from a vector resonance.
It was pointed out in Ref.~\cite{Angelescu:2015uiz,Gupta:2015zzs,Altmannshofer:2015xfo} that in scenario like
2-Higgs Doublet Model (2HDM), including the Minimal Supersymmetric Standard Model (MSSM) and the Next-to-Minimal Supersymmetric
Standard Model (NMSSM), the branching ratio $Br(H/A\rightarrow \gamma\gamma)$ turns out to be very small ${\cal O}(10^{-6})$. It was further
noted that it remains small even in the extreme case of $\tan\beta\sim 1$ which is the lower limit required by the Renormalization Group Equation (RGE) running of Yukawa couplings. In a recent paper~\cite{Ding:2015rxx}, $R$-parity SUSY model is considered
to address the diphoton excess issue.

In this paper, we propose an simple extension of 
 the SM with an additional doublet Higgs $H^{'}$ and a singlet $s$. We consider the
 neutral component $H^{\prime }_0$ of $H'$ as the 750 GeV resonance, and assume that
 $s$ is lighter than 2.6 GeV. Especially,  $H^{\prime }_0$ can be produced at the LHC 
 at tree level via $q{\bar q}$ production, and decay into a pair of $s$ at tree level.
 And then $s$ can decay into a pair of collimated photons. Because $s$ is light and 
 highly boosted, each pair of photons may appear as an single photon in the detector~\cite{Chala:2015cev,Agrawal:2015dbf}.
  We show that the diphoton production cross section can be from
 3 to 13 ${\rm fb}$, the decay width of $H^{\prime }_0$ can be from 30 to
 60 GeV, and all the current experimental constraints including dijet constraint
 can be satisfied.

This paper is organized as follows. We devote Section~\ref{sec:2} to describe our model. In Section~\ref{sec:3}, we investigate the diphoton signal in our model by considering constraints from the LHC at  $\sqrt{s}=$ 8 TeV. 
We also discuss the scenario where two collimated photon pairs coming from $s\rightarrow \gamma \gamma$ can be measured as two photons and 
can be accounted for the diphoton excess. In addition to it, we discuss two scenarios in which we can enhance 
$Br(s\rightarrow \gamma \gamma)$. Conclusion and summary are given in Section~\ref{sec:4}.

\begin{figure}[htp!]
\centering
\subfiguretopcaptrue

\subfigure{
\includegraphics[totalheight=5.5cm,width=12.cm]{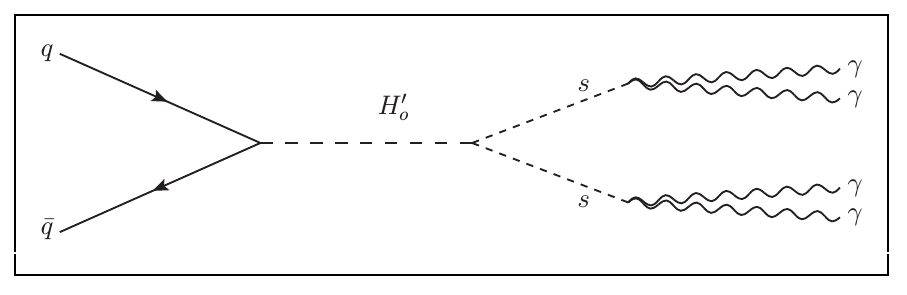}
}

\caption{Feynman diagram for the process $q\bar q\rightarrow H_{0}^{'}\rightarrow s s \rightarrow \gamma \gamma \gamma \gamma$.
}
\label{fig1}
\end{figure}

\section{The Model Building}\label{sec:2}
The model we propose here is a simple extension of the SM. This model contains two Higgs doublets, $H$ and $H^{'}$ where $H$ is a
SM like Higgs doublet, as well as a SM real singlet $s$. For simplicity, we assume that $H^{'}$ will not develop a vacuum expectation value ($VEV$) or its VEV is negligible,
and $s$ is CP-even. The discussion for CP-odd singlet is similar.
The Lagrangian of the model can be given as:
\begin{align}
{\mathcal{L}} =&[-m_{H}^2|H|^{2} + m_{H^{'}}^2|H^{'}|^2 + \frac{1}{2} m_{s}^2 s^{2}+ \lambda_{1}|H|^{4} 
+\lambda_{2}|H^{'}|^{4} + \lambda_{3}|H|^{2}|\bar{H^{'}}|^{2} +\lambda_{4}|{\bar H^{'}}H|^{2}] \nonumber \\
&-[\frac{\lambda_{5}}{2}({\bar H^{'}}H)^{2}+ Y_{d}Hq{\bar d}+Y_{u}{\bar H}q{\bar u}+Y_{e}Hl{\bar e} 
+{\xi_{1}}{\bar H^{'}}q{\bar d}+{\xi_{2}}H^{'}q{\bar u} +\frac{\lambda_{s}}{2}H^{'}Hs^{2}+.... +h.c]
\label{eq:leg}
\end{align}
\noindent where  $m_{H}$, $m_{H^{'}}$, $m_{s}$ are the masses of  SM like Higgs $H$, the extra Higgs $H^{'}$ and the scalar $s$, $\lambda_{1}$ is the crucial quartic 
coupling which is fixed by the Higgs mass, $\lambda_{2}-\lambda_{5}$ couplings are not relevant in our present study and we set them to zero for simplicity, while $Y_{u,d}$ are the conventional Yukawa couplings. Furthermore, we also assume that  $\xi ={\xi_{1}}={\xi_{2}}$ which are 
the additional couplings between the extra Higgs and quarks. Here we want to emphasize that since in our case quarks directly couple to $H^{'}$, the production rate of resonance is 
($p p \rightarrow H^{'}$) is very large as compared to the conventional loop induced production, such as the gluon fusion. 
This is one of the distinct feature of our scenario as compared to \cite{Chala:2015cev}. In the last part of Eq.~\ref{eq:leg}, $\lambda_{s}$ represents the strength of the interaction of $H^{'}$, $H$ and $s$. This term plays the role in determining the decay

\section{A Promising Mechanism to Generate Diphoton Excess}\label{sec:3}
In this section we will elaborate the mechanism we follow to address the issue of diphoton excess.

\subsection{Diphoton Excess and the LHC Constraints}\label{sec:3.1}
\begin{figure}[htp!]
\centering
\subfiguretopcaptrue
\subfigure{
\includegraphics[totalheight=6.5cm,width=8.cm]{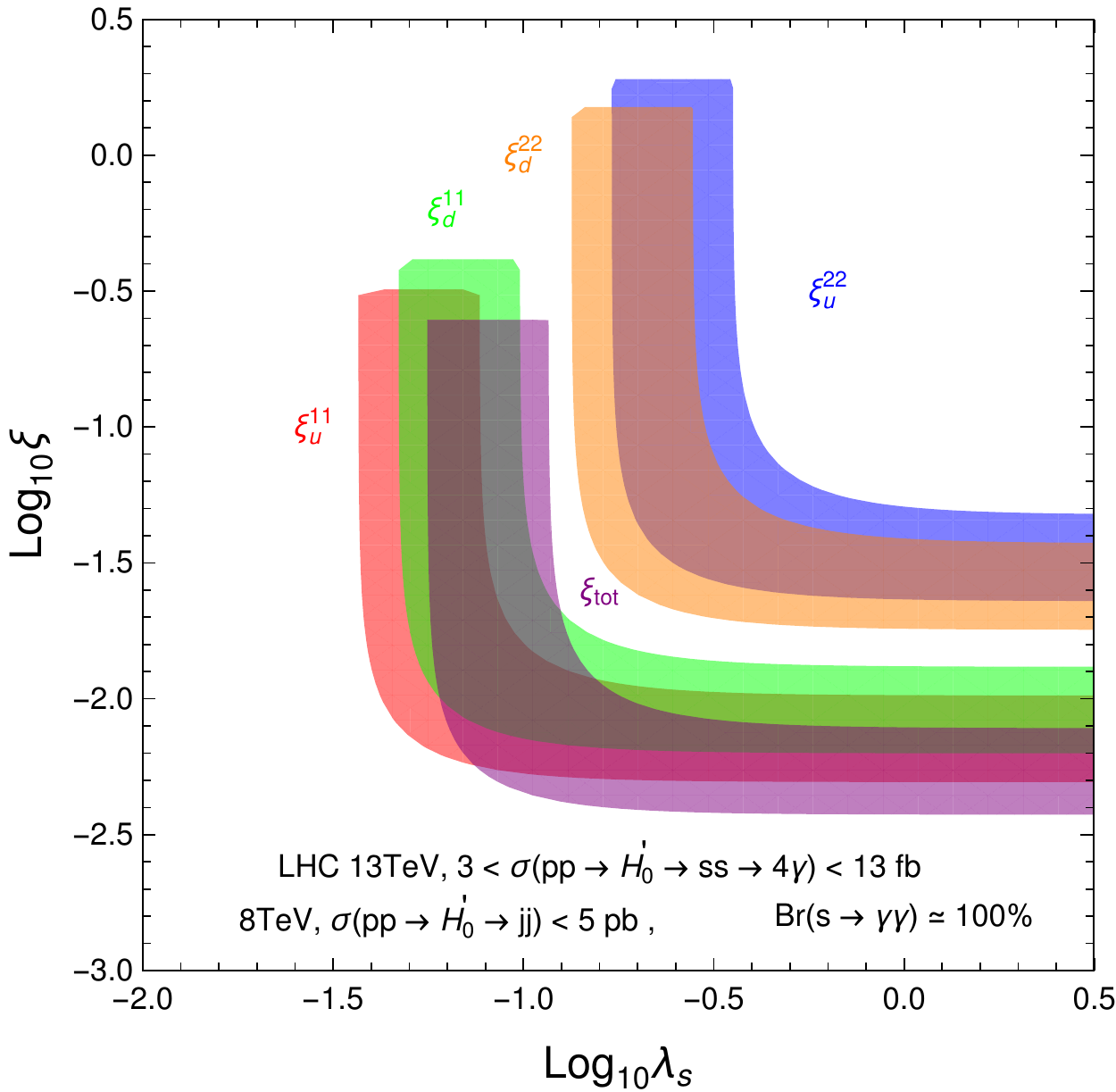}
}
\subfigure{
\includegraphics[totalheight=6.5cm,width=8.cm]{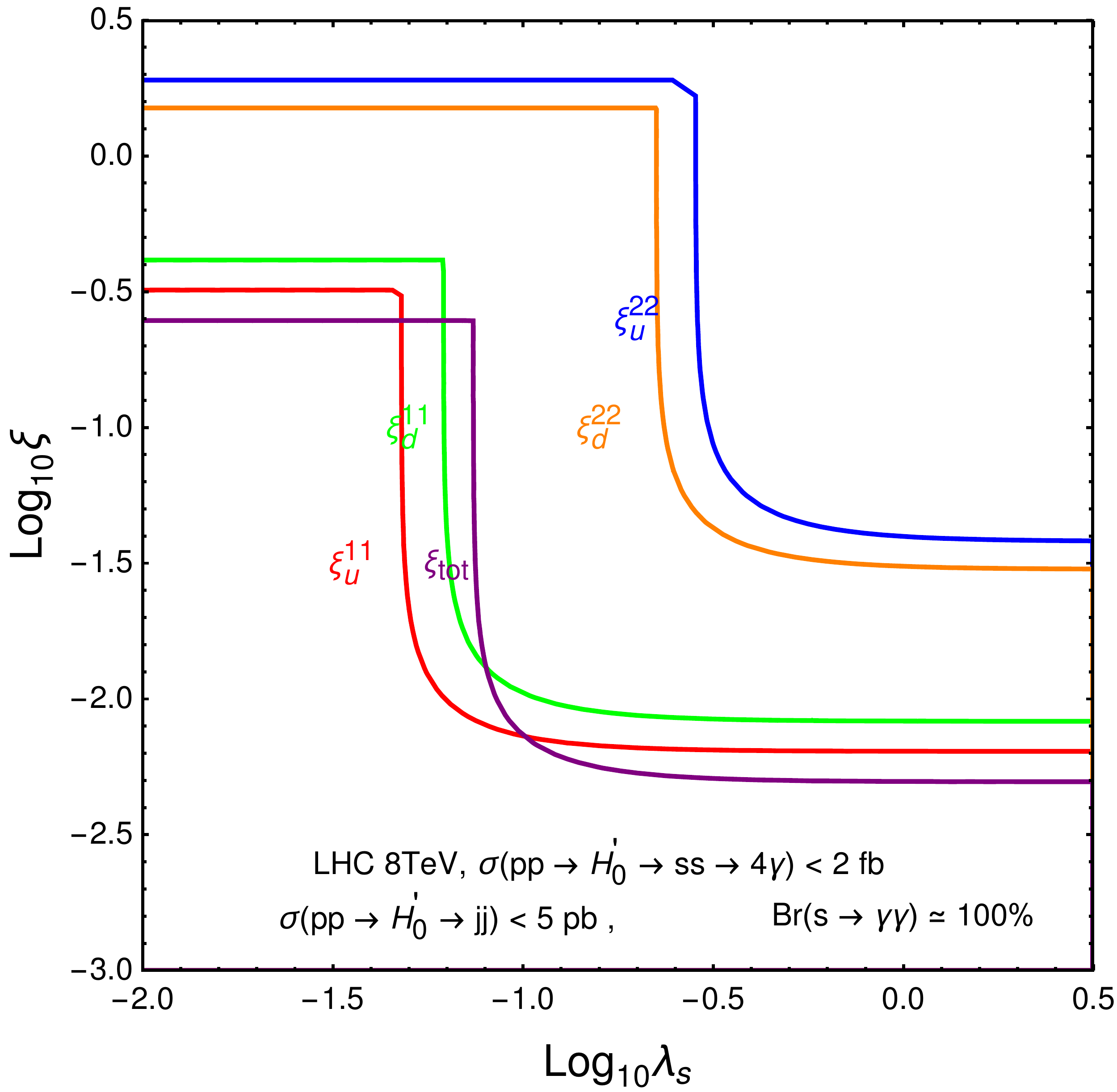}
}

\caption{Left panel: allowed region for of diphoton rate on the $[\log_{10}\lambda_s,~\log_{10}\xi]$ plane, we take  the diphoton cross section
$3{\rm fb}<\sigma^{13}_{pp\to\gamma\gamma}<13$~fb 
and dijet constraint $\sigma^{8}_{pp\to H_0^{'}\to jj} <5$~Pb. The red, green, blue, orange and purple colors are corresponding to $\xi_u^{11}$, $\xi_d^{11}$, $\xi_u^{22}$, $\xi_d^{22}$ and $\xi_{\rm tot}$ contributions, respectively. 
Right panel: similar plots for the LHC 8 TeV, where the upper bound of diphoton rate is set to be $2$ fb. 
 }
\label{signal}
\end{figure}

\begin{figure}[htp!]
\centering
\subfiguretopcaptrue
\subfigure{
\includegraphics[totalheight=10.0cm,width=10.cm]{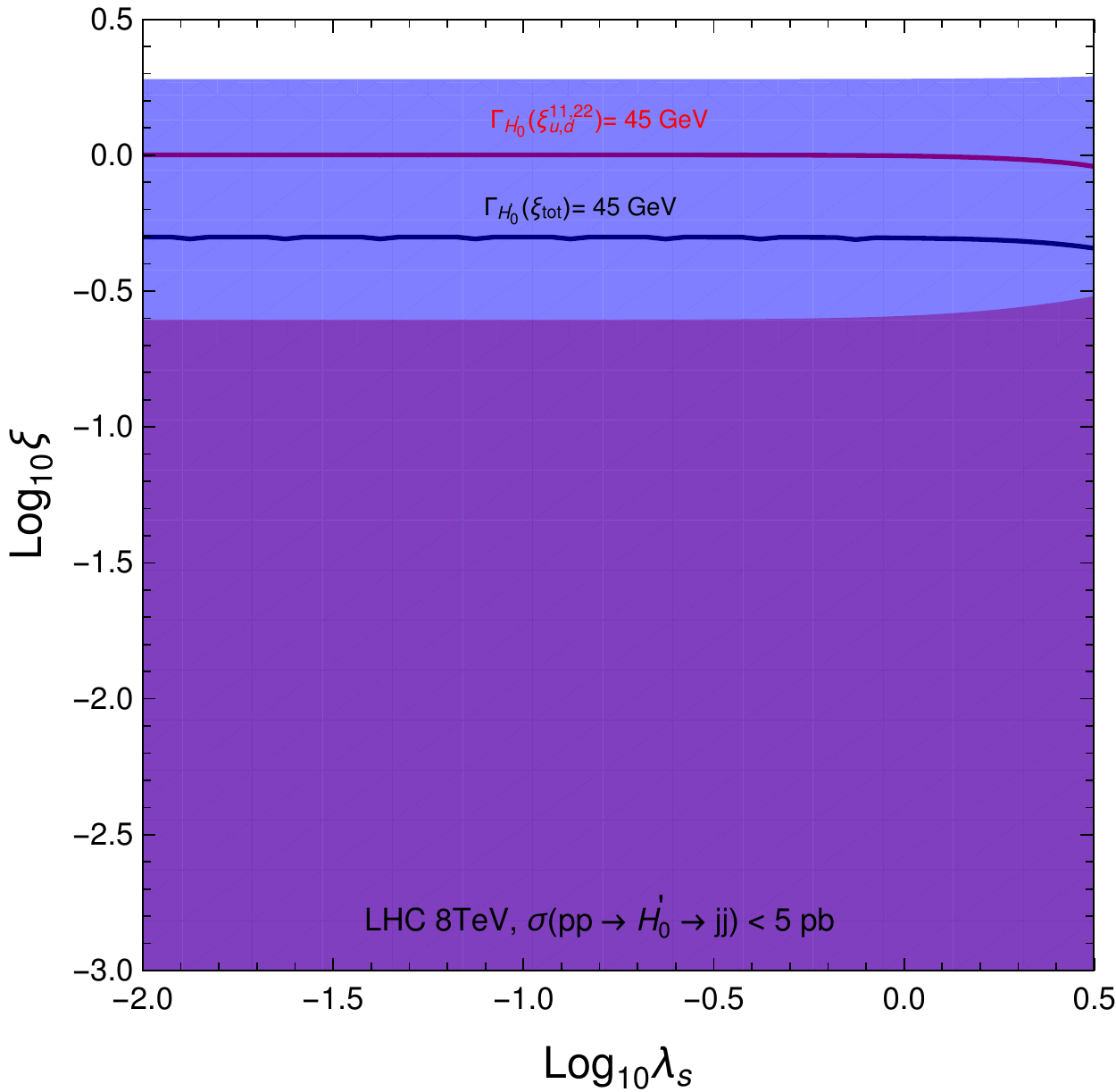}
}

\caption{ 
The contours of total decay widths $\Gamma_{H^{\prime 0}} (\xi_{u,d}^{11,22})=45$~GeV (upper line)
and $\Gamma_{H^{\prime 0}} (\xi_{\rm tot})=45$~GeV (lower line) are presented in 
the $[\log_{10}\lambda_s,~\log_{10}\xi]$ plane.
For comparison, the allowed regions for the dijet constraints  are also 
given with the same color legends in Fig.~\ref{signal} for $\xi_{u}^{22}$ 
and $\xi_{\rm tot}$, respectively.
}
\label{Viable-Region}
\end{figure}

To estimate the diphoton signal quantitatively, we use {\tt SARAH}~\cite{sarah} generate {\tt UFO} model file~\cite{Degrande:2011ua}, and {\tt MadGraph5\_aMC@NLO}~\cite{MG5} to calculate the production cross section of $H_{0}^{'}$ with   {\tt CTEQ6L1}~\cite{Nadolsky:2008zw} parton distribution function (PDF). 
We consider  the diphoton cross section
$3{\rm fb}<\sigma^{13}_{pp\to\gamma\gamma}<13$~fb 
and dijet constraint $\sigma^{8}_{pp\to H_0^{'}\to jj} <5$~Pb.
We display our results of calculations in $\lambda_{s}-\xi$ plane. Plot in the left panel is
for LHC 13 TeV. In this plot red, green, blue, orange and purple colors are corresponding to $\xi_u^{11}$, $\xi_d^{11}$, $\xi_u^{22}$, $\xi_d^{22}$ and $\xi_{\rm tot}$ contributions, respectively. Here, $\xi_{\rm tot}$ is the universal coupling for
the first two generations.
In the right panel we present the similar plot for LHC 8 TeV where the upper bound of diphoton rate is set to be $2$ fb. 
It should be noted that for couplings of first generation quarks $\xi_u^{11}$ and $\xi_d^{11}$, the viable overlapped region on the diphoton cross section
and dijet constraint is smaller than that for the second generation quarks
 $\xi_u^{22}$ and $\xi_d^{22}$. Therefore, if one combines the results of diphoton excess at $13$ TeV and null result at $8$ TeV, $H_{0}^{'}$ coupling to second generation quarks are more favored. In Fig.~\ref{Viable-Region}, we show 
 the contours of total decay widths $\Gamma_{H^{\prime 0}} (\xi_{u,d}^{11,22})=45$~GeV 
and $\Gamma_{H^{\prime 0}} (\xi_{\rm tot})=45$~GeV  in 
the $[\log_{10}\lambda_s,~\log_{10}\xi]$ plane.
For comparison, the allowed regions for the dijet constraints  are also 
given with the same color legends in Fig.~\ref{signal} for $\xi_{u}^{22}$ 
and $\xi_{\rm tot}$, respectively.
Therefore, the case with $\xi_{\rm tot}$ is entirely excluded by the dijet
constraints. Interestingly, for the case with $\xi_{u,d}^{22}$, the dijet constraint
can be evaded due to the PDF dependence.
In short,  the diphoton production cross section can be from
 3 to 13 ${\rm fb}$, the decay width of $H_0^{\prime }$ can be from 30 to
 60 GeV, and all the current experimental constraints including dijet constraint
 can be satisfied.

\subsection{Collimated Photons from Boosted Scalars}\label{sec:3.2}
In this subsection we will discuss how two pairs of photons generated from $s\rightarrow \gamma \gamma$ may be falsely detected as diphoton  
instead of four photons. Similar analysis is also given in Refs.~\cite{Chala:2015cev,Agrawal:2015dbf}. The main idea behind this assumption is that in decaying 
process, $s\rightarrow \gamma \gamma$, if the $s$ is highly boosted, so that it travels longer before decaying and the pair of photons are highly collimated, then it is possible that the two photons when arrive at the detector may be falsely detected as a single photon. It is 
important to note that $s$ can not be very long lived otherwise it will decay outside electromagnetic calorimeter (ECAL) or can not be very
short lived. In that case pair of photons will not be too collimated and four gamma events will be registered instead of diphoton.
 Let us describe this scenario in more detail. The distribution of opening angle 
$\alpha$ between the collimated photons in the Lab frame is given as:
 
\begin{align}
\frac{dN}{d\alpha}=\frac{{\cos\frac{\alpha}{2}}} {2{\gamma}{\beta} {\sin^{2}{\alpha/2}}\sqrt{{\gamma^2}\sin^{2}{\alpha/2}-1}},
\label{open-ang}
\end{align}   
\noindent where $\gamma$ and $\beta$ are the boost factor and velocity of $s$ and range of $\alpha$ is $[2 \sin^{-1}{\frac{1}{\gamma}}, \pi]$. 
For $\gamma >>$ 1, $\alpha_{min}=2 \sin^{-1}{\frac{1}{\gamma}}\sim \frac{2}{\gamma}$. 
We find a difference of factor 2 in the denominator of Eq.~(\ref{open-ang}) as compared to Eq.~(3.1) of Ref.~\cite{Chala:2015cev}. Integrating Eq.~(\ref{open-ang})
we get:
\begin{align}
N(\alpha)=\frac{\sqrt{-2+\gamma^{2}-\gamma^{2}\cos{\alpha}}\csc{\frac{\alpha}{2}}}{\sqrt{2}\beta\gamma}.
\end{align} 
\noindent We note that for $\alpha_{mp} \approx \frac{4.6}{\gamma}$, $N(\alpha_{mp}) =$ 0.9, that is for about 90$\%$ of the scalar decaying to a pair
of photons, the opening angle is $\alpha \lesssim \frac{4.6}{\gamma}$. It is noted that the CMS ECAL has a resolution of
$\Delta \eta \times \Delta \phi =0.0174 \times 0.0174$ and has radius R= 1.3 meters~\cite{Khachatryan:2015iwa} while for ATLAS has a resolution of 
$\Delta \eta \times \Delta \phi =0.025 \times 0.025$ and has radius R=1.5 meters~\cite{Aad:2009wy,Aad:2010sp}. 
If we assume the $s$ particle flies transversely and decays to two photon. Then, the photons reach to ECAL at a distance: 
\begin{align}
\Delta z =\alpha_{mp}(R-\beta\gamma \hat{c\tau})
\label{del-z}
\end{align}
\noindent $\hat{c\tau}$ is the proper lifetime of a particular scalar $s$. Note that $\beta\gamma \hat{c\tau}$ is just
a decay length of $s$ can also be given as:
\begin{align} 
l=&\frac{\beta\gamma}{\Gamma}
\label{decay:len}
\end{align}
\noindent where $\Gamma$ is the decay width of $s$.
Since we are assuming for simplicity that $s$ decays 
perpendicular to the beam line implies $\Delta z/R < \Delta \eta$. For our case we can write Eq.~(\ref{del-z}) as 
\begin{align}
\Delta \eta\approx& \alpha_{mp}(1-\frac{\beta\gamma \hat{c\tau}}{R}) \nonumber \\
 =&\frac{4.6 m_{s}}{375} - \frac{4.6}{R\Gamma} 
\label{del-eta}
\end{align}
\noindent Here we use the fact $\alpha_{mp}=4.6/\gamma=4.6 m_{s}/375$ and mass $m_{s}$ is mass of $s$.
From Eqs.~(\ref{del-z}) and (\ref{del-eta}), we can deduce the following two bounds
\begin{align}
1-\frac{\gamma \hat{c\tau}}{R}=1 - \frac{375 \hat{c\tau}}{m_{s}R}=1 - \frac{375}{m_{s}\Gamma R } > 0, \\
\Delta \eta_{s} \approx \frac{4.6 m_{s}}{375}- \frac{4.6}{\Gamma R} \le \Delta \eta .
\label{eta:bounds}
\end{align}

\begin{figure}[htp!]
\centering
\subfiguretopcaptrue

\subfigure{
\includegraphics[totalheight=6.5cm,width=10.5cm]{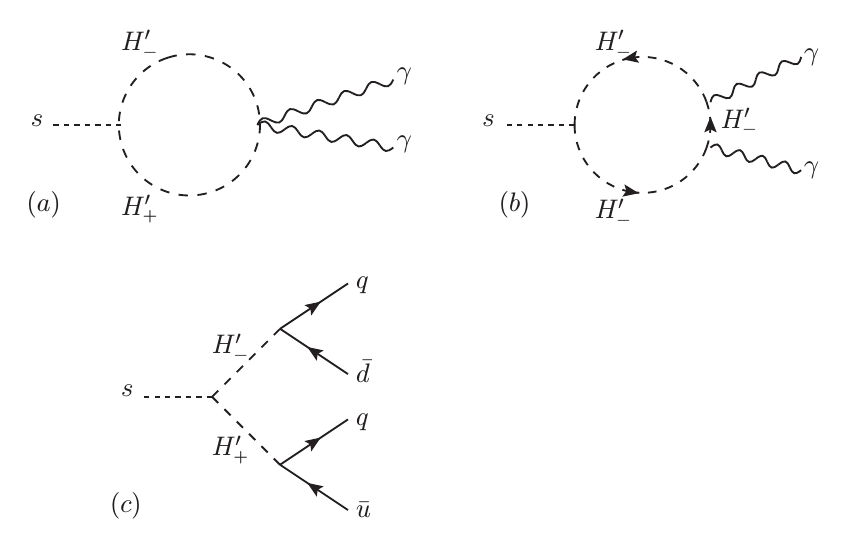}
}

\caption{Feynman diagram for scenario ($I$).
}
\label{feyn2}
\end{figure}

\begin{figure}[htp!]
\centering
\subfiguretopcaptrue

\subfigure{
\includegraphics[totalheight=4.5cm,width=7.5cm]{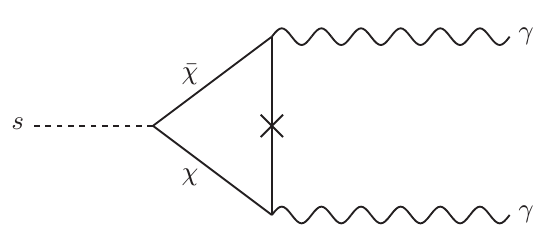}
}

\caption{Feynman diagram for scenario ($II$).
}
\label{feyn3}
\end{figure}


\begin{figure}[htp!]
\centering
\subfiguretopcaptrue
\subfigure{
\includegraphics[totalheight=6.5cm,width=8.cm]{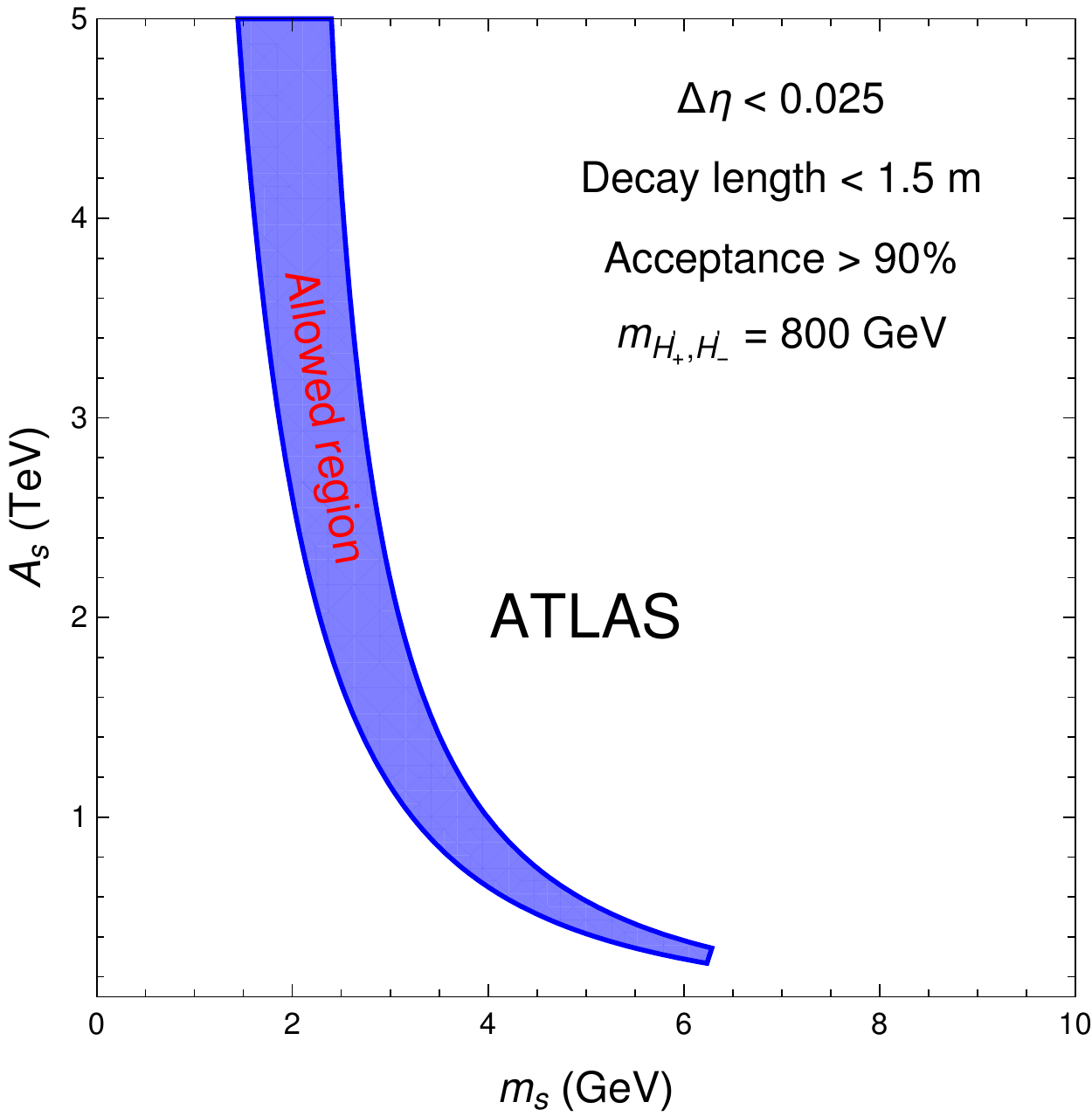}
}
\subfigure{
\includegraphics[totalheight=6.5cm,width=8.cm]{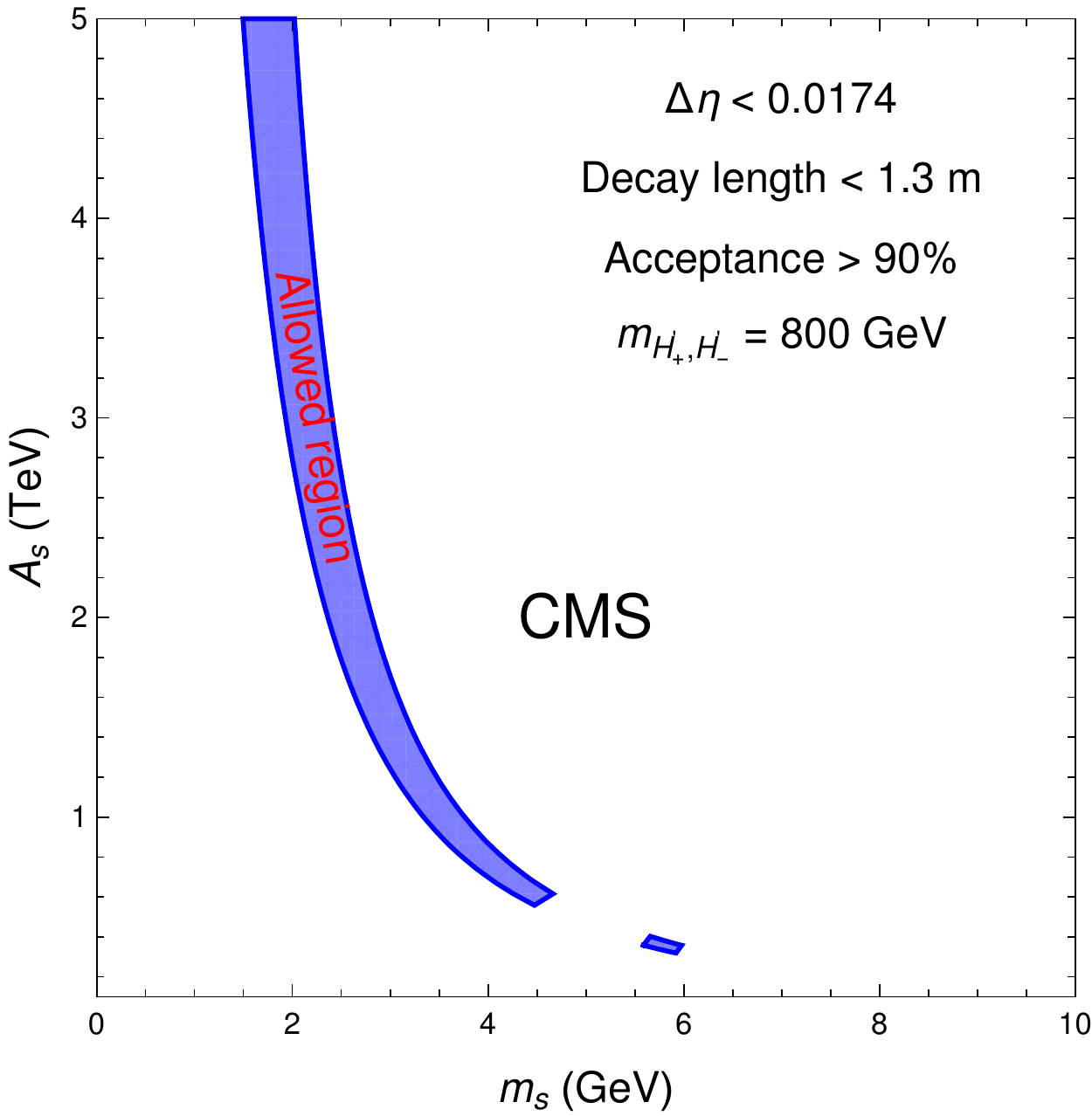}
}

\caption{
Plots in $m_{s}-A_{s}$ plane. Blue region represents allowed parameter space.}
\label{scenario:1}
\end{figure}

\subsection{Enhancement of $Br(s\rightarrow \gamma \gamma)$}\label{sec:3.3}

After discussing how two pairs of photons generated from a pair of decaying scalar may disguise themselves as diphoton, we will discuss two scenarios through which we can enhance $Br(s\rightarrow \gamma \gamma)$.

{\bf Scenario (I)}: In this scenario we propose to add a following term to Eq.~(\ref{eq:leg})
\begin{align}
{\mathcal{L}} =&A_{s}s|H^{'}|^{2}
\label{sce-1}
\end{align}
\noindent where $A_{s}$ is a trilinear scalar coupling. The corresponding Feynman diagrams are shown in \ref{feyn2}. In Fig.~\ref{feyn2}a,b we see that $s$ decays to diphoton via charged Higgs loops, while Fig.~\ref{feyn2}c depicts that $s$ can also decay to quarks respectively.
Here we argue that in Fig.~\ref{feyn2}a, $g_{EM}$ gauge coupling is involved as compared to Fig.~\ref{feyn2}b which is suppressed
because of $g_{EM}^2$. In Fig.~\ref{feyn2}c, we see that because of off-shell charged Higgs, this diagram in general is suppressed. It should be 
noted that here we are considering only the second generation of quarks. Moreover, if we restrict the parameter space by imposing the kinematical constrain $m_{s} < 2 m_{c}$, then the diphoton final state is the only available channel in which $s$ can decay.

We display results of our calculations
in $m_{s}-A_{s}$ plane in Fig.~\ref{scenario:1}. In these plots we set $m_{H_{\pm}^{'}}=$ 800 GeV. 
In the left panel, we also use ATLAS ECAL specifications (that is radius $R$ and $\Delta \eta$ values) related to our work as discussed in Section~\ref{sec:3.2}
and the allowed parameter space is shown in blue color.
In this plot, we use Eqs.~(\ref{decay:len}) and (\ref{eta:bounds}) to calculate decay length $l$ of $s$ and estimate $\Delta \eta$ and restrict ourselves to $l < 1.5$ meters and $\Delta \eta <$ 0.025. Here, we see that for relatively large values of $m_{s} \sim$ 6 GeV, 
the allowed $A_s \sim$ 400 GeV. But as the $m_{s}$ decreases, $A_{s}$ starts rising up and at around $m_{s}\sim$ 2.5 GeV, the allowed parameter
space becomes essentially insensitive of $A_{s}$ values. But we do note that the width of allowed band somewhat increases for small values 
of $m_{s}$. We present similar analysis in the right panel with the CMS ECAL specifications for $R$ and $\Delta \eta$. Here we restrict the parameter space by imposing  $l < 1.3$ meters and $\Delta \eta <$ 0.0174. We immediately note that with slightly tight constraints, the allowed region of
parameter space shrinks in width. It can also be seen that at about $m_{s}\sim$ 2.5 GeV, the allowed parameter
space becomes essentially insensitive of $A_{s}$ values.  We note that the allowed mass range for ${s}$ consistent
with all the current constraints discussed above is about $m_{s}\sim$ [1.4,~6] GeV and [1.5,~4.5] GeV for the ATLAS and CMS specifications, respectively.

\begin{figure}[htp!]
\centering
\subfiguretopcaptrue
\subfigure{
\includegraphics[totalheight=6.5cm,width=8.cm]{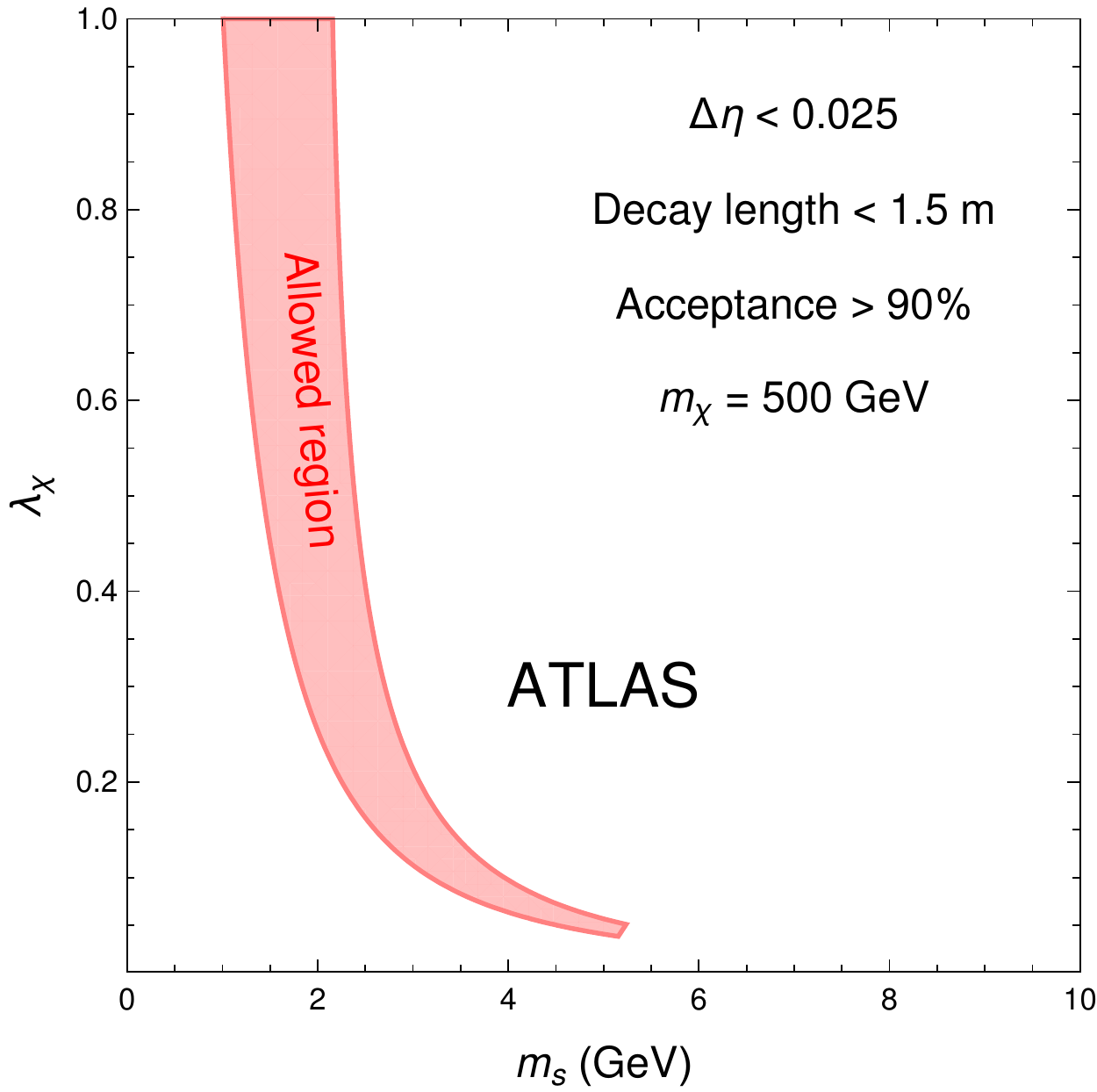}
}
\subfigure{
\includegraphics[totalheight=6.5cm,width=8.cm]{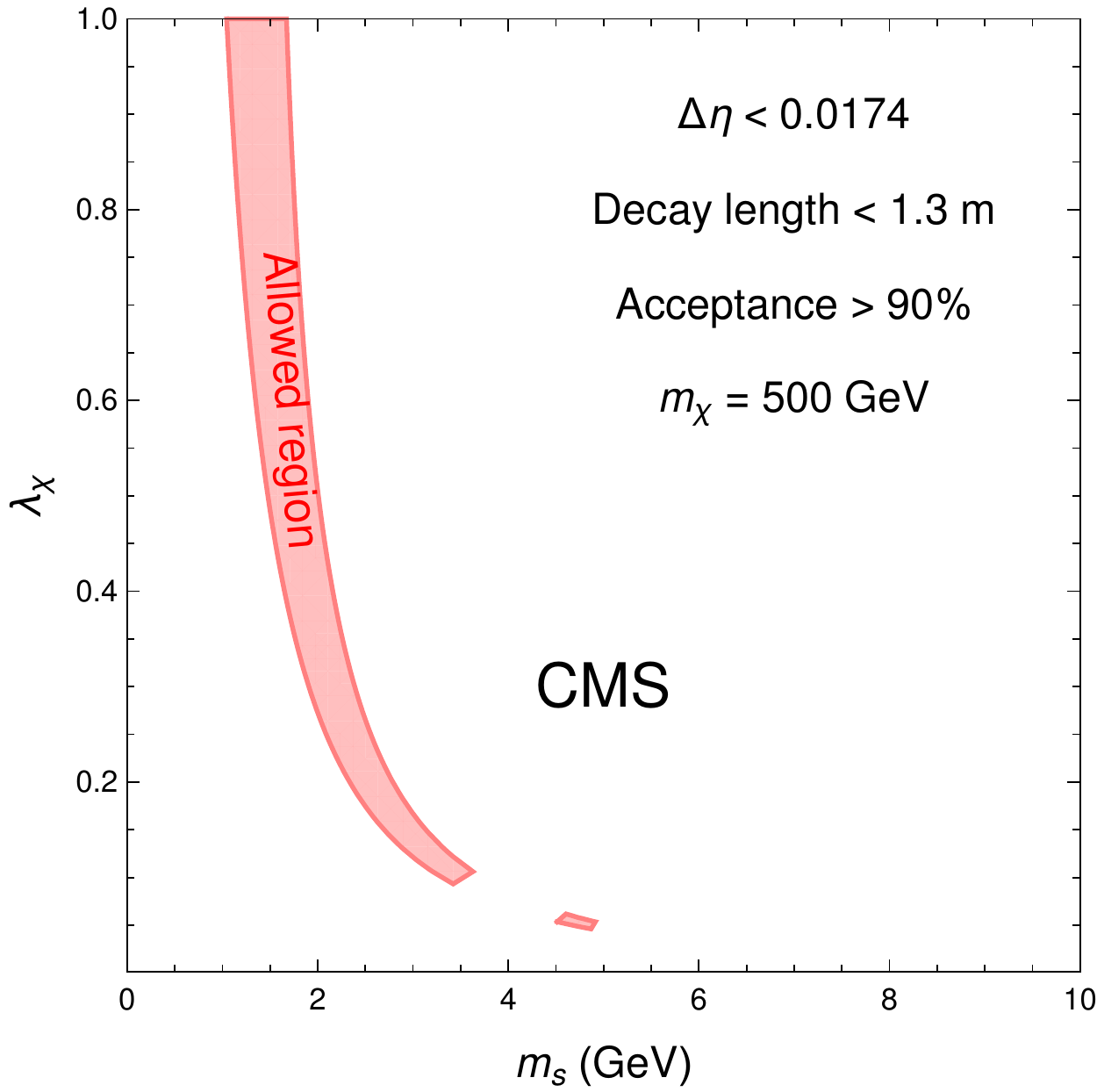}
}

\caption{
Plots in $m_{s}-\lambda_{\chi}$ plane. Blue region represents allowed parameter space. }
\label{scenario:2}
\end{figure}

{\bf Scenario (II)}: In this scenario we propose to add the vector-like fermions
 ($\chi$, ${\bar\chi}$) with electromagnetic charges $\pm1$
 like the right-handed charged leptons to our model by introducing the following term to Eq.~(\ref{eq:leg})
\begin{align}
{\mathcal{L}} =&\lambda_{\chi}s{\chi} {\bar\chi} + m_{\chi}{\chi} {\bar\chi}
\label{sce-1}
\end{align}
\noindent The corresponding Feynman diagram is shown in Fig.~\ref{feyn3}. Note that here we only add vector-like leptons (VLL) $\chi$, since
addition of vector-like quarks (VLQ) can generate gluon pairs and will form jets. They are highly constrained by the dijet searches.
In Fig.~\ref{scenario:2}, we presents our calculations in $\lambda_{\chi}-m_{s}$ plane. For these plots, we fix $m_{\chi} =$ 500 GeV.
Like Fig.~\ref{scenario:1}, plot in the left panel follows the ATLAS ECAL specifications for $R$ and $\Delta \eta$ and the plot in the right panel satisfies the CMS ECAL
specifications and allowed parameter space is shown in red color band. 
The allowed region starts slightly widening and becomes independent of $\lambda_{\chi}$ values as 
$m_{\chi}$ moves towards smaller values. Similar behaviour can also be seen in the right panel but with somewhat narrow allowed region.
We note in scenario (II), as compared to scenario (I), the maximum value of $m_{s}$ is about 5 GeV for $\lambda_{\chi}\sim$ 0.05. 
We see that the allowed mass range for $m_{s}$ consistent with all the constrains we discussed above,
$m_{s}\sim$ [1,~5] GeV and [1,~3.5] GeV for ATLAS and CMS specifications, respectively


\section{Summary and Conclusion}\label{sec:4}

An excess of events in diphoton channel with invariant mass of about 750 GeV have been reported by the ATLAS and CMS Collaborations. Considering it as a tantalizing hint for 
 new physics beyond the SM, we proposed a SM extension 
  with an additional doublet Higgs $H^{'}$ and a singlet $s$. We considered the
 neutral component $H^{\prime }_0$ of $H'$ as the 750 GeV resonance, and assumed that
 $s$ is lighter than 2.6 GeV. In particular,  $H^{\prime }_0$ can be produced 
 at tree level via $q{\bar q}$ production, and decay into a pair of $s$ at tree level.
 And then $s$ can decay into a pair of collimated photons.
 Because $s$ is highly boosted and appropriately long lived, the pair of photons coming from a decaying scalar can
be collimated enough to be measured as a single photon event in the detector. Thus,
 one can understand the excess in the diphoton events
as a result of four photon final states instead of two photons.
  We showed that the diphoton production cross section can be from
 3 to 13 ${\rm fb}$, the decay width of $H^{\prime }_0$ can be from 30 to
 60 GeV, and all the current experimental constraints including dijet constraint
 can be satisfied.

{\bf Acknowledgements}--
This research was supported in part by the Natural Science Foundation of China
under grant numbers 11135003, 11275246, 11475238 (TL) and 11475191, 11135009 (XJB).


\end{document}